\documentclass[%
aps,prf,superscriptaddress,amsmath,amssymb,floatfix,preprint,showpacs,showkeys]{revtex4-1}

\usepackage{graphicx}
\usepackage{bm}
\usepackage{mathtools}
\usepackage{xcolor}

\begin{document}

\title{A fast and efficient deep learning procedure for tracking droplet motion in dense microfluidic emulsions}


\author{Mihir Durve}
\affiliation{Center for Life Nano Science@La Sapienza, Istituto Italiano di Tecnologia, Viale Regina Elena, 291, 00161 Roma, Italy}
\affiliation{Quantitative Life Sciences Unit, The Abdus Salam International Centre for Theoretical Physics (ICTP), Trieste 34151, Italy}

\author{Fabio Bonaccorso}
\affiliation{Center for Life Nano Science@La Sapienza, Istituto Italiano di Tecnologia, Viale Regina Elena, 291, 00161 Roma, Italy}
\affiliation{Istituto per le Applicazioni del Calcolo CNR, via dei Taurini 19, Rome, Italy}
\affiliation{Department of Physics and INFN, University of Rome Tor Vergata, Via della Ricerca Scientifica 1, 00133, Rome, Italy}

\author{Andrea Montessori}
\affiliation{Center for Life Nano Science@La Sapienza, Istituto Italiano di Tecnologia, Viale Regina Elena, 291, 00161 Roma, Italy}
\email{a.montessori@iac.cnr.it}

\author{Marco Lauricella}
\affiliation{Center for Life Nano Science@La Sapienza, Istituto Italiano di Tecnologia, Viale Regina Elena, 291, 00161 Roma, Italy}

\author{Adriano Tiribocchi}
\affiliation{Center for Life Nano Science@La Sapienza, Istituto Italiano di Tecnologia, Viale Regina Elena, 291, 00161 Roma, Italy}
\affiliation{Istituto per le Applicazioni del Calcolo CNR, via dei Taurini 19, Rome, Italy}

\author{Sauro Succi}
\affiliation{Center for Life Nano Science@La Sapienza, Istituto Italiano di Tecnologia, Viale Regina Elena, 291, 00161 Roma, Italy}
\affiliation{Istituto per le Applicazioni del Calcolo CNR, via dei Taurini 19, Rome, Italy}
\affiliation{Institute for Applied Computational Science, John A. Paulson School of Engineering and Applied Sciences, Harvard University, Cambridge, USA}

\date{\today}






\begin{abstract}
We present a deep learning-based object detection and object tracking
algorithm to study droplet motion in dense microfluidic emulsions.
The deep learning procedure is shown to correctly predict the droplets' shape and track their motion
at competitive rates as compared to standard clustering algorithms, even in the presence of significant deformations. 
The deep learning technique and tool developed in this work could be used for the general study of 
the dynamics of biological agents in fluid systems, such as moving cells and 
self-propelled micro organisms in complex biological flows.
\end{abstract}

\keywords{Deep learning, YOLO and DeepSORT, Object recognition and tracking, Lattice Boltzmann approach, Dense emulsions.}

\maketitle

\section{Introduction}

In recent times, Machine Learning (ML) has gained enormous attention within the 
scientific community \cite{LeCunn}, as it provides a general automated framework to accomplish highly complex computational tasks through a self-improving trial and error minimization procedure ~\cite{mitchell}.

Although often overhailed (for a critical assessment see  \cite{Coveney, SucciCoveney}), ML
proves indeed successful in a number of tasks, such as object classification and image recognition 
~\cite{wu,rawat}, handwriting recognition~\cite{darmatasia}, text analysis~\cite{hasan,ramadhani}, voice 
recognition~\cite{tandel}, and more recently also physical \cite{CARLEO} and biological 
applications such as protein folding \cite{Alfa_Fold}.

Machine learning algorithms perform these tasks by realising a model that outputs
decisions or predictions starting from (big) data inputs which are used to train the 
network by recursive optimisation of its parameters (the weights connecting nodes
of the network across two subsequent layers). A particularly successful instance of ML, called deep learning (DL), uses 
multi-layer artificial neural networks,  as opposed to the so-called shallow networks, which 
consist of just a few hidden layers ~\cite{goodfellow}.
Deep learning-based approaches have been successfully employed in a number of applications, from medical diagnosis ~\cite{Chen2016,guo2019,fujiyoshi2019,xie2019} to physics\cite{sharir2020deep,CARLEO}. In microfluidics, deep learning has been used to study physical systems and quality control processes. Mahdi et al.~\cite{mahdi} used neural networks to predict the size of the droplets in an emulsion. Khor et al ~\cite{khor} developed a convolutional autoencoder model to discover a low-dimensional representation to describe droplet shapes within a concentrated emulsion and study the stability of droplets in an emulsion. Hadikhani et al ~\cite{hadikhani} trained neural networks to estimate fluid and flow parameters by analysing digital images of fluid in microchips.

\begin{figure}
  \centering
  \includegraphics[scale =0.80]{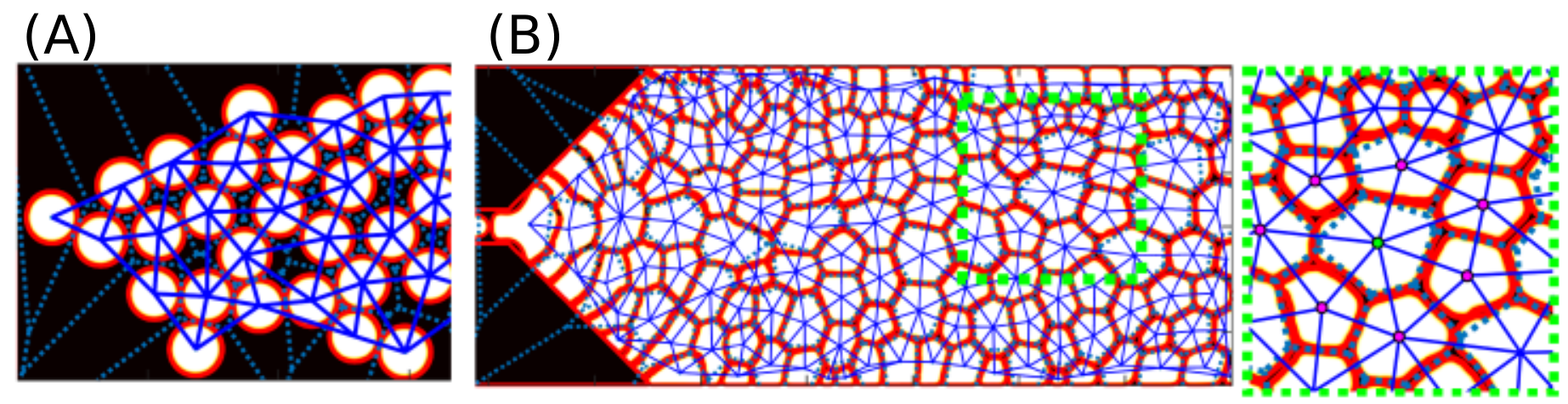}
  \caption{\label{figLB} LB simulation of (A) Early stage formation of soft flowing crystals in microchannel. (B) Flowing high internal phase emulsion in microfluidic channel with divergent inlet. Blue solid lines connecting centers of droplets result from a Delaunay triangulation, while dotted polygons are the associated Voronoi tessellation.}
\end{figure}

In recent years, high internal phase emulsions (HIPE) and soft flowing crystals, namely ordered states of flowing matter characterised by a highly ordered collection of liquid droplets arranged in crystal-like  structures, are showing their potential in many fields of science and engineering. Indeed, due to the typical porous polymeric matrix (see figure \ref{figLB} for two examples of emulsions in microfluidic channels (a) during the early stage formation and (b) at high packing fractions ), they can be efficiently employed in a wide range of advanced applications, such as catalyst supports, ion-exchange modules, separation media, electrochemical sensing, to name but a few. 
Moreover, they are very fit for bio-inspired applications such as scaffolds in tissue engineering, where the interconnected pore structure is essential for their function. 
Besides their relevance for a host of applications in the fields mentioned above, soft porous
materials also raise a fundamental challenge to non-equilibrium thermodynamics, since they display properties which cannot be traced back to any of the three basic states of matter. For instance, HIPEs are binary mixtures of immiscible liquids, each featuring linear Newtonian  rheology, that exhibit highly non-Newtonian mechanical and rheological properties.
Thus, a full-understanding of the rich dynamics underlying these classical many-body systems is required to 
optimise the microfluidic devices employed for their production and to fine-tune their properties for different purposes.

The aim of this work is to demonstrate that general deep learning-based tools can 
be adapted, tuned and profitably used to extract trajectories of droplets 
starting from digital frames of the fluid system, as obtained either by 
experiment or via computer simulations.
The deep learning procedure is here applied to the case of dense flowing emulsions 
in micro-channels to compute the paths followed by the droplets, as obtained via Lattice Boltzmann simulations. This represents a very first step towards a more complex task, consisting in the characterization of the dynamics of this many-body soft system in terms of effective equations of motion of the droplets flowing in the continuum phase. 

The results are encouraging, in that the procedure proves capable of fast (about 400 frames/second) recognition and tracking of droplets with significant departures from the spherical shape.
Given its flexibility, it is expected that the procedure developed in this paper can be readily generalized to the study of the dynamics of moving agents in micro and nanoflows of biological interest~\cite{Bernaschi_melchionna_succi_RMP2019}.

The paper is structured as follows. In the next section we describe the Machine Learning procedure and, shortly, the lattice Boltzmann method used to produce the validation data. Section~\ref{section_results} is dedicated to a discussion 
of the numerical results and some conclusive remarks close the paper.

\section{Methods}

In the following we provide details of the Machine Learning procedure as well as  
of the Lattice Boltzmann method, which is the simulation technique employed to produce the validation data.  We begin by discussing the latter.

\subsection{Lattice Boltzmann method for dense emulsions}
\label{sec:LB}

In this section we briefly describe the numerical model, namely an extended color-gradient approach with repulsive near-contact interactions, previously employed in Ref. \cite{montessori2019mesoscale,montessori2019modeling,Montessori2021prfluids,montessori2021softmatter,montessori2020philtransa}. In the multicomponent LB model, two sets of discrete distribution functions evolve via  the usual streaming-collision algorithm (see \cite{succi2018lattice,kruger2017lattice}):

\begin{equation} \label{CGLBE}
f_{i}^{k} \left(\vec{x}+\vec{c}_{i}\Delta t,\,t+\Delta t\right) =f_{i}^{k}\left(\vec{x},\,t\right)+\Omega_{i}^{k}[ f_{i}^{k}\left(\vec{x},\,t\right)] +S_i(\vec{x},t),
\end{equation}

In the above equation, $f_{i}^{k}$ is the discrete distribution function, representing the probability of finding a particle of the $k-th$ component at position $\vec{x}$, time $t$ with discrete velocity $\vec{c}_{i}$, and $S_i$ is a source term coding for the effect of external forces (such as gravity, near-contact interactions, etc). 

In equation \ref{CGLBE} the time step is taken as $\Delta t= 1$, and the index $i$ runs over the discrete lattice
directions $i = 1,...,b$, being $b=9$ for a two  dimensional nine speed lattice (D2Q9).
The density $\rho^{k}$ of the $k-th$ component and the total linear momentum of the mixture 
$\rho \vec{u}=\sum_k\rho^{k}\vec{u}^k $  are obtained, respectively, via the zeroth and the first order moment of the lattice distributions
$\rho^{k}\left(\vec{x},\,t\right) = \sum_i f_{i}^{k}\left(\vec{x},\,t\right)$ and 
$\rho \vec{u} = \sum_i  \sum_k f_{i}^{k}\left(\vec{x},\,t\right) \vec{c}_{i}
$.
The collision operator splits into three parts \cite{gunstensen1991lattice,leclaire2012numerical}: 

\begin{equation}
\Omega_{i}^{k} = \left(\Omega_{i}^{k}\right)^{(3)}\left[\left(\Omega_{i}^{k}\right)^{(1)}+\left(\Omega_{i}^{k}\right)^{(2)}\right].
\end{equation}

where $\left(\Omega_{i}^{k}\right)^{(1)}$, stands for the standard collisional relaxation \cite{succi2018lattice}, 
$\left(\Omega_{i}^{k}\right)^{(2)}$ codes for the perturbation step \cite{gunstensen1991lattice}, contributing to the buildup of the interfacial tension while $\left(\Omega_{i}^{k}\right)^{(3)}$ is the recoloring step \cite{gunstensen1991lattice,latva2005diffusion}, which promotes the segregation between the two species, minimising their mutual diffusion.

By performing a Chapman-Enskog multiscale expansion it can be shown that the hydrodynamic limit of Eq.\ref{CGLBE} is a
set of equations for the conservation of mass and linear momentum (i.e., the Navier-Stokes equations), with a capillary stress tensor of the form:
\begin{equation}\label{capstress}
\bm{\Sigma}=-\tau\sum_i \sum_k\left(\Omega_{i}^{k}\right)^{(2)} \vec{c}_i \vec{c_i}= \frac{\sigma}{2 |\nabla \rho|}(|\nabla \rho|^2\bm{I} - \nabla \rho \otimes \nabla \rho)
\end{equation}
being $\tau$ the collision relaxation time, controlling the kinematic viscosity via the relation  
$\nu=c_s^2(\tau-1/2)$ ( $c_s=1/\sqrt{3}$ the sound speed of the model) and $\sigma$ is the surface 
tension \cite{succi2018lattice,kruger2017lattice}. 

In eq. \ref{capstress}, the symbol $\otimes$ denotes a dyadic tensor product. 
The stress-jump condition across a fluid interface is then augmented with an intra-component repulsive term aimed at 
condensing the effect of all the repulsive near-contact forces (i.e., Van der Waals, electrostatic, steric 
and hydration repulsion) acting on much smaller scales ($\sim  O(1 \; nm)$)  than those resolved 
on the lattice (typically well above hundreds of nanometers).
It takes the following form:
\begin{equation}
\bm{T}^1\cdot \vec{n} - \bm{T}^2 \cdot \vec{n}=-\nabla(\sigma \bm{I} - \sigma (\vec{n}\otimes \vec{n})) - \pi \vec{n}
\end{equation}
where $\pi[h(\vec{x})]$ is responsible for the repulsion between neighbouring fluid interfaces, 
$h(\vec{x})$ being the distance between interacting interfaces along the normal $\vec{n}$.\\


The additional term can be readily included within the LB framework, by adding a 
forcing term acting only on the fluid interfaces in near-contact, namely:
\begin{equation}
\vec{F}_{rep}= - A_{h}[h(\vec{x})]\vec{n} \delta_I
\end{equation}

In the above, $A_h[ h(\vec{x})]$ is the parameter controlling the strength (force per unit volume)
of the near-contact interactions, $h(\vec{x})$ is the distance between the interfaces, $\vec{n}$ is a unit vector normal to the interface and $\delta_I\propto \nabla\phi$ is a function, proportional to the phase field $\phi=\frac{\rho^1-\rho^2}{\rho^1+\rho^2}$, employed to localize the force at the interface.

To conclude, the addition of the repulsive force (added to the right hand side of Eq.~\ref{CGLBE}) naturally leads to the following (extended) 
conservation law for the momentum equation:
\begin{equation} \label{NSEmod}
\frac{\partial \rho \vec{u}}{\partial t} + \nabla \cdot {\rho \vec{u}\vec{u}}=-\nabla p + \nabla \cdot [\rho \nu (\nabla \vec{u} + \nabla \vec{u}^T)] + \nabla \cdot (\bm{\Sigma}  +   \pi \bm{I})
\end{equation}

namely  the Navier-Stokes equation for a multicomponent system, augmented with a surface-localized repulsive 
term, expressed through the gradient of the potential function $ \pi$.

\subsection{Deep learning procedure}

\begin{figure}[h]
  \centering
  \includegraphics[scale=0.8]{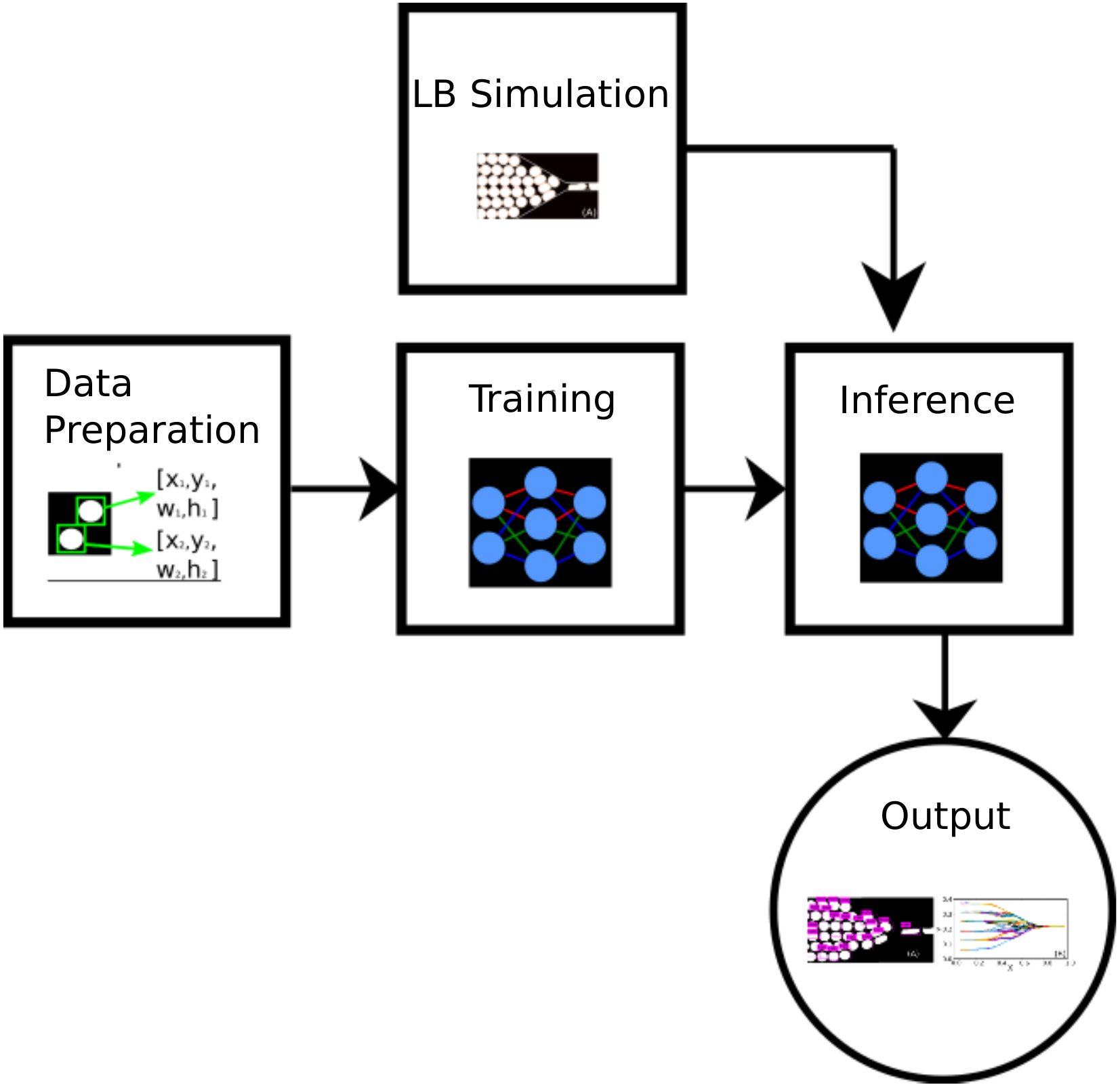}
  \caption{\label{illustration} The three basic steps involved in the deep learning based application 
  for trajectory extraction of droplets in dense emulsions. From left to right: 
  1) Synthetic data preparation for the network training, 2) Network Training, 3) Inference and Validation
  based on the data provided by the LB simulation (top).  The output is a set of droplets along
  with their trajectories (bottom).
  }
\end{figure}

The Deep learning approach employed in this work consists of three basic steps: 
1) Data preparation, 2) Network training and set-up, and 3) Inference (see Fig.~\ref{illustration}), which we now discuss in some detail.

In order to extract trajectories of droplets from the digital images, the deep 
learning-based application must accomplish two tasks: i) identify the droplets and find their location in a given image, and ii) track the droplet displacements between successive frames. In our implementation, these two tasks are performed via two different algorithms:
droplet detection makes use of YOLO \cite{redmon,redmon1},  a state-of-the-art, deep neural network algorithm for real-time object detection. The YOLO algorithm has been shown to be significantly faster than most other techniques for object detection and classification~\cite{redmon1,sanchez}. The YOLO algorithm is highly accurate in detecting objects on interest in the images consisting of many objects and complex background scenarios. YOLO algorithm can be implemented on the CPU as well as GPUs to gain higher rate of image analysis (inference speed) making it a good choice for real-time video analysis. YOLO takes an image as an input and outputs a list of detected objects with their classes, confidence of detection, and their locations. This information is then passed to DeepSORT \cite{wojke}, an effective multi-object tracker, whose task is precisely to keep track of the detected objects as they move in time.

\subsubsection{Data preparation}

In order to identify droplets in a given image, the YOLO network must be trained with several examples, called training data set, which should have a wide enough statistical significance. Its preparation is arguably the most labour intensive task of the training phase and will have high impact on the validity of the whole procedure. For supervised learning algorithms, training data typically consist of inputs, such as images, and expected output of the network called {\it labels}. In order to train a network to identify droplets in an image, the training data must include several examples, each consisting of an image of a system with droplets, along with a text file containing their sizes (height and width) and centers positions (x and y coordinates) within the image as labels. 

\begin{figure}[h]
  \centering
  \includegraphics[width=15cm,height=15cm,keepaspectratio]{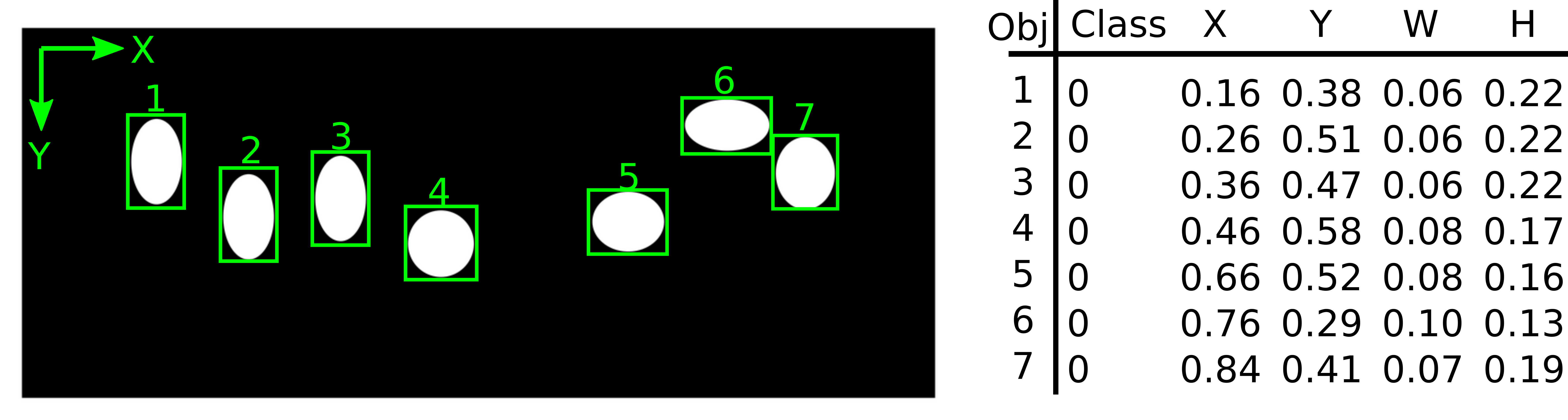}
  \caption{\label{data_label} A sample image from training data set with its label in the format needed to train YOLO network, namely the sequential number of the object, the class of the object (droplet), the coordinates of its center (x and y), the width (w) and height (h) of the bounding box.}
\end{figure}

To generate the training data set, we prepared several images by randomly placing a fixed number (7) of droplets with varying aspect ratios in a uniformly dark background of size 1040 X 450 pixels, since this is the typical output of the LB simulations. For each sample, we build the image and note the positions and sizes of all the droplets that we placed. An example from the training data set is shown in Fig.~\ref{data_label}, along with the text file containing the labels of each droplet, namely the coordinate of its center and the height and width of its bounding box. The code to generate the training data is given in the data access sheet.

\subsubsection{Network training and set-up}

The network set-up is performed by using Darknet \cite{darknet}, an open-source neural network framework written in C and CUDA and developed to support YOLO on both CPU and GPU devices: it supports the most common neural network operations such as convolution, max pooling, and various activation functions. Once the network is trained, the same framework can then be used as an inference engine.

\begin{figure}[h]
  \centering
  \includegraphics[width= 15cm, height=15cm,keepaspectratio]{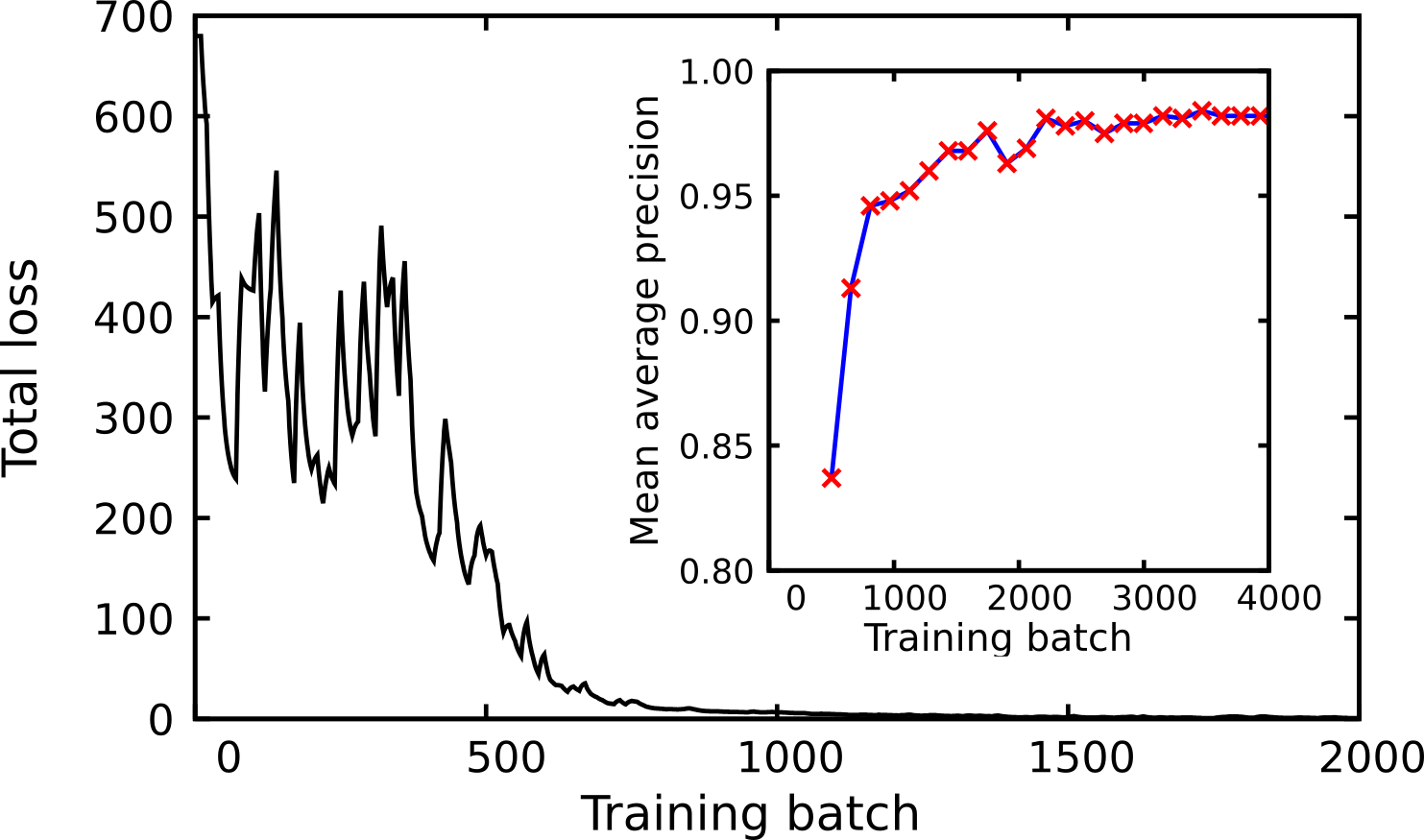}
  \caption{\label{Fig_loss} Total loss function as the training progresses. 
 In the inset we show mean average precision on the validation data computed during various stages of the training.}
\end{figure}

For our task, we selected a predefined, ready to implement network architecture, known as yolov3-tiny, consisting of {\it 24 layers}, branching out in many convolutional and pooling layers with a total of approximately {\it 8 million} trainable parameters (weights).
The training of the YOLO network was carried out by employing {\it $10,000$} training images as described above. For the training, we used GPU implementation of Darknet on a system consisting of NVIDIA RTX 2060 Super GPU, common in middle-level desktop configurations for gaming. The Github repository outlines the details to carry out training of the network~\cite{darknet_repo1} as well as in the data access sheet. The training operation is realized in many batches. In each batch, a fixed number of images 
(called batch size) are given as input to the network. Based on the predicted output of the network, a loss function is calculated, and the parameters of the network are then updated to minimize it. The YOLO algorithm uses a multi-component loss function~\cite{redmon,redmon1} accounting for the errors in classifying detected objects, locating the objects, as well as error in prediction confidence for the detected objects. The chosen values of parameters for the training (batches, batch size, etc.) can be found in ~\cite{para_list}. As the training progresses, the total loss decreases to a near-zero value, as shown in Fig. \ref{Fig_loss}. 

The accuracy of a trained network for object detection is measured by a metric called 
Average Precision (AP),  a figure of merit between 0 and 1. 
The AP of a network can be computed using a separate validation data set and in our case, 
we reserved a fraction (5\%) of the training data set as validation data set. In order to compute the AP, all the images of the validation data set are given as input to the network and all detected objects (droplets) are tabulated in descending order of their individual detection confidence score. Starting from the top of the table, for each detected object, precision and recall values are computed as follows.

\begin{align}
P = \frac{TP}{TP + FP}, && R= \frac{TP}{TP+FN}.
\end{align}

Here, TP is the number of true positive detection (droplet does exist and it is detected), FP is 
the number of false positive detection (i.e. a droplet is detected but it doesn't exist ) and FN 
is the number of false negative detection (i.e., a droplet does exist but it is not detected). 

\begin{figure}[h]
  \centering
  \includegraphics[width= 10cm, height=10cm,keepaspectratio]{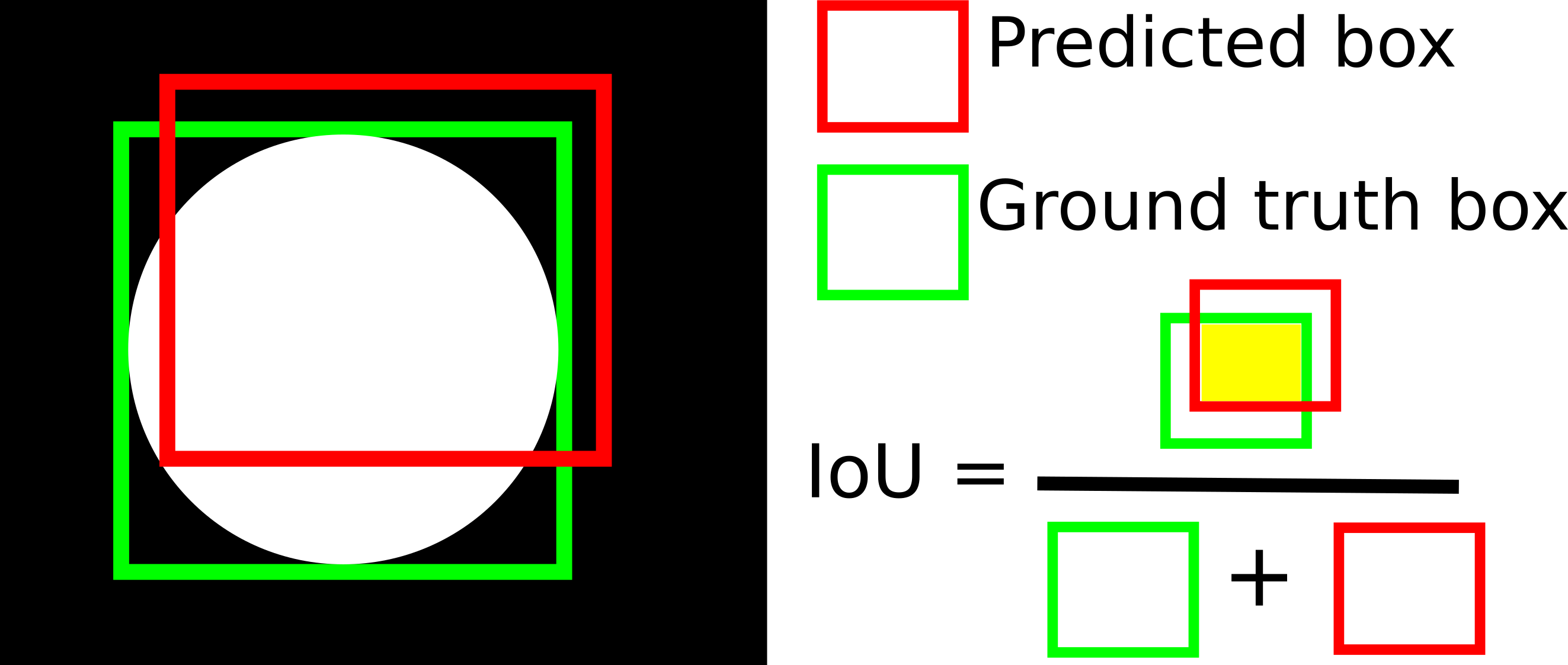}
  \caption{\label{iou} Intersection over union of prediction (red box) and ground truth (green box).}
\end{figure}

To determine whether the detection is a true positive or a false positive, we use a quantity called 
{\it Intersection over Union (IoU)} value of the detection. 
In geometric terms, IoU is given by the area of intersection divided by area of union of ground truth bounding box, 
with predicted bounding box (see Fig.~\ref{iou}). We set the IoU threshold to $0.5$, i.e. if the IoU is $\ge 0.5$ the prediction is classified as a true positive.

After computing Precision and Recall for all the predictions in the validation data set, the Precision-Recall curve is plotted
and the AP is computed as, $AP = \int_{0}^{1} p(r) \,dr$. 
We also measured the mean average precision (mAP), defined as the average precision among 
all the classes of objects in the data set, as reported in the inset of Fig. \ref{Fig_loss}. 
We observe that the mAP value monotonically increases as the training progresses and 
saturates to about $0.98$ after a sufficient number of training batches, indicating that the
training procedure achieves nearly perfect accuracy. At the end of the training, the optimised parameters of the network 
are saved and used for the Inference step, which we discuss next.

\subsubsection{Inference step}

In the Inference step, the input data, i.e. the density maps produced by the LB simulations at each time step, are presented as an input to the trained network so that YOLO detects the droplets in the simulated image and passes this information along to the object tracking algorithm (DeepSORT). 
Together, YOLO and DeepSORT detect and track the droplets at each time step, producing all the trajectories.

The particular network architecture can produce accurate trajectories even if the dimension in pixel of the input image is not exactly the same used during the training phase: training is a slow process that scales with the image size. It is a good advantage to reuse the coefficients of the Yolo network for images even 2x or 3x bigger, for which the training would have been significantly slower. 

In the next section, we apply the network to two microfluidic cases simulated using the Lattice Boltzmann method described in Sec.~\ref{sec:LB}, namely a straight and tapered channel, respectively.
The first scenario should be easy for the detection (droplets are almost always circles) and tracking (droplets do not merge or split). The second case is a more interesting one, since the nozzle forces the droplets to aggregate and squeeze to fit the central channel: the shape of each droplet changes in time and some break-ups are present, which makes it a good test case.

\section{Results and discussion}
\label{section_results}

\begin{figure*}[h]
  \centering
  \includegraphics[width=\textwidth,height=5cm,keepaspectratio]{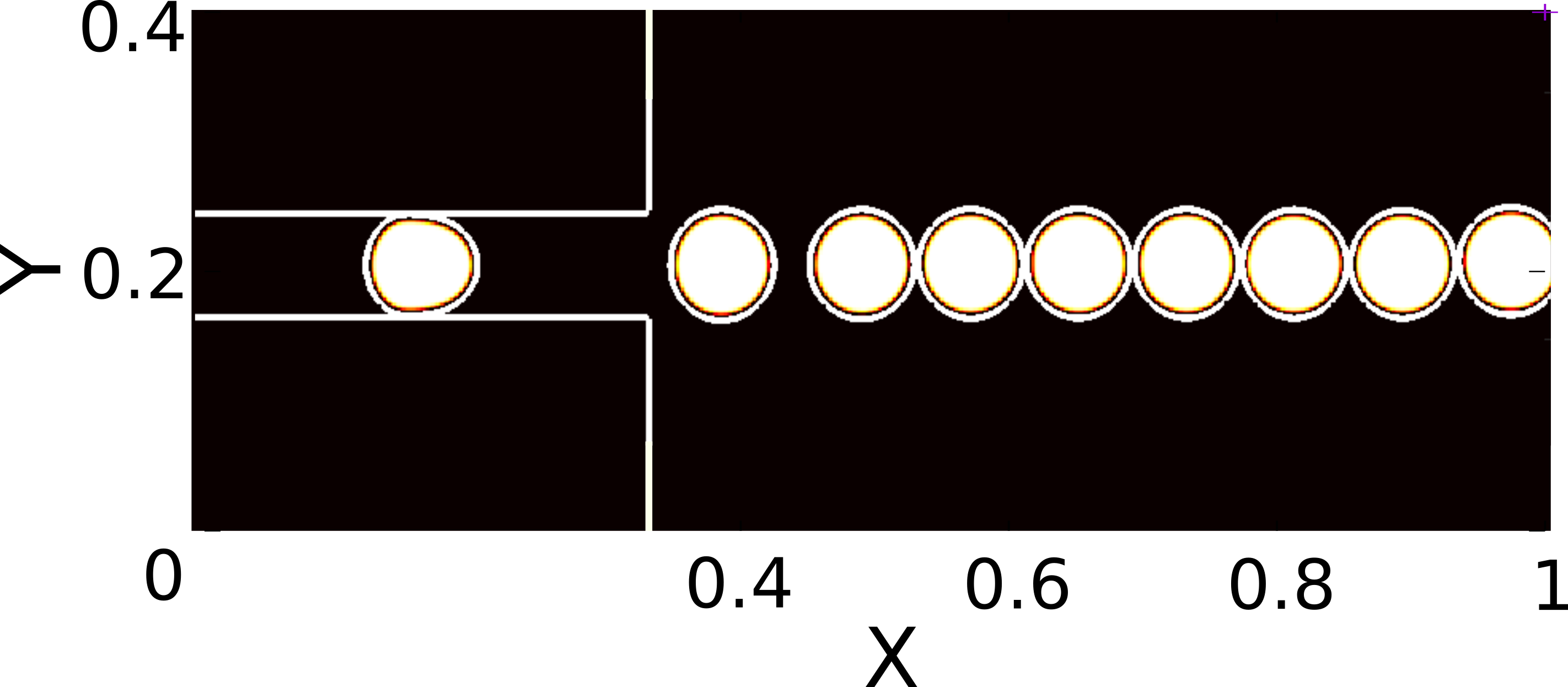}
  \caption{\label{Fig_simple} Snapshot of the Simple channel. Droplets flow from left to right, in a almost straight horizontal path after leaving the channel at the left. The motion is quasi one-dimensional and leaves the droplets nearly circular.}
  \end{figure*}

To test our trained network we analysed video data generated for the two aforementioned applications. For the details, see the data access sheet. The first case (\textbf{Simple channel}) consists of a thin inlet channel in which droplets are continuously produced and flow downstream in the main microchannel (see Fig.~\ref{Fig_simple}). Sequences of images of size 850 X 250 pixels are generated and stacked to prepare a video file which serves as an input to the trained YOLO network for droplet detection. 

\begin{figure*}[h]
  \centering
  \includegraphics[width=\textwidth,height=5cm,keepaspectratio]{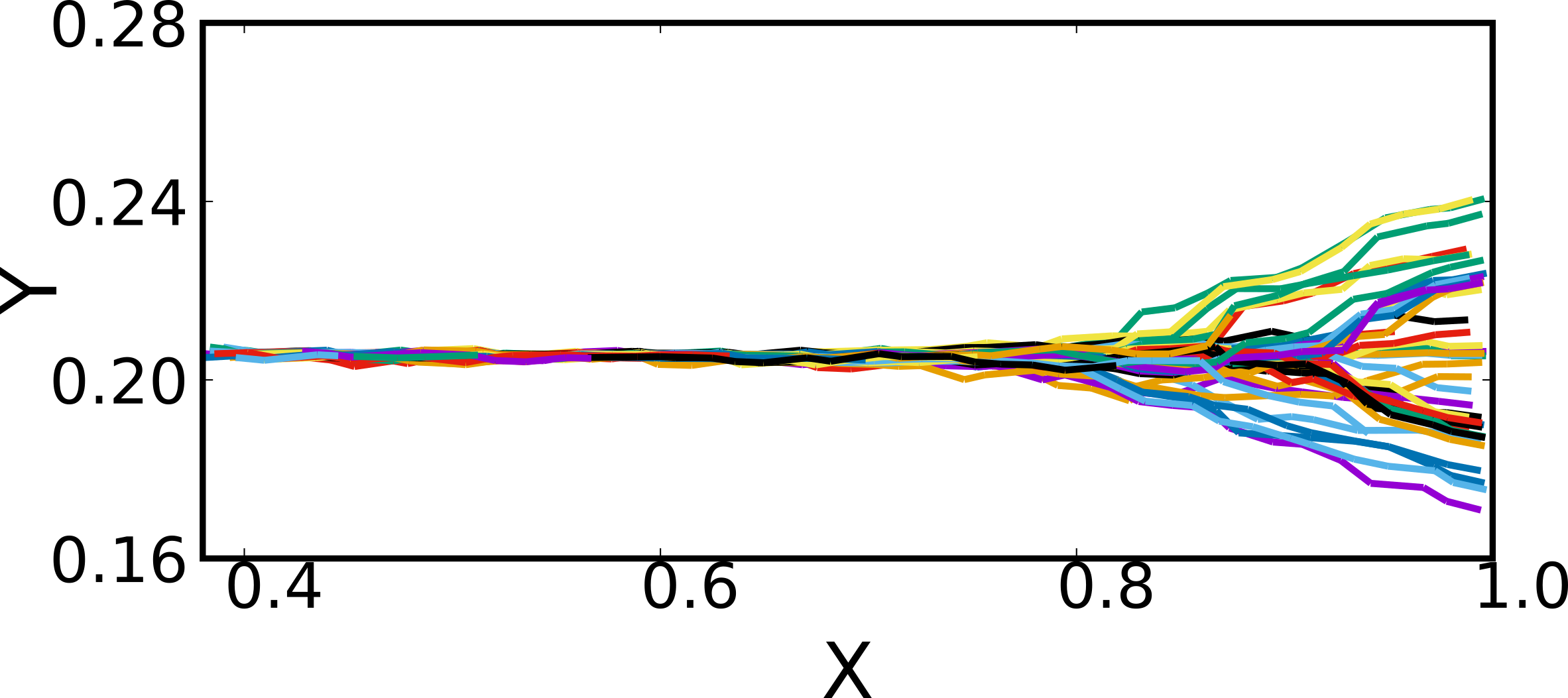}
  \caption{\label{traj_simple} Trajectories (centers of the bounding boxes) of the droplets for the Simple channel. Along the Y-axis small variations are correctly detected far enough from the channel. Each colour represents the trajectory of an individual droplet.  }
\end{figure*}

The YOLO network analyses each frame of  the video file, computing the bounding boxes for each detected droplet. The information of the bounding boxes, i.e. their size and location, is then passed to DeepSORT to track the droplets in the sequential images using a unique ID associated to each detected droplet.
In Fig.~\ref{traj_simple} the inferred trajectories, as the centers of the bounding boxes, of the tracked droplets are shown.
A video file (simple\_channel.mp4) showing the detection and tracking of the droplets is provided in the Supplementary Material. 

\begin{figure*}[h]
  \centering
  \includegraphics[width=\textwidth,height=5cm,keepaspectratio]{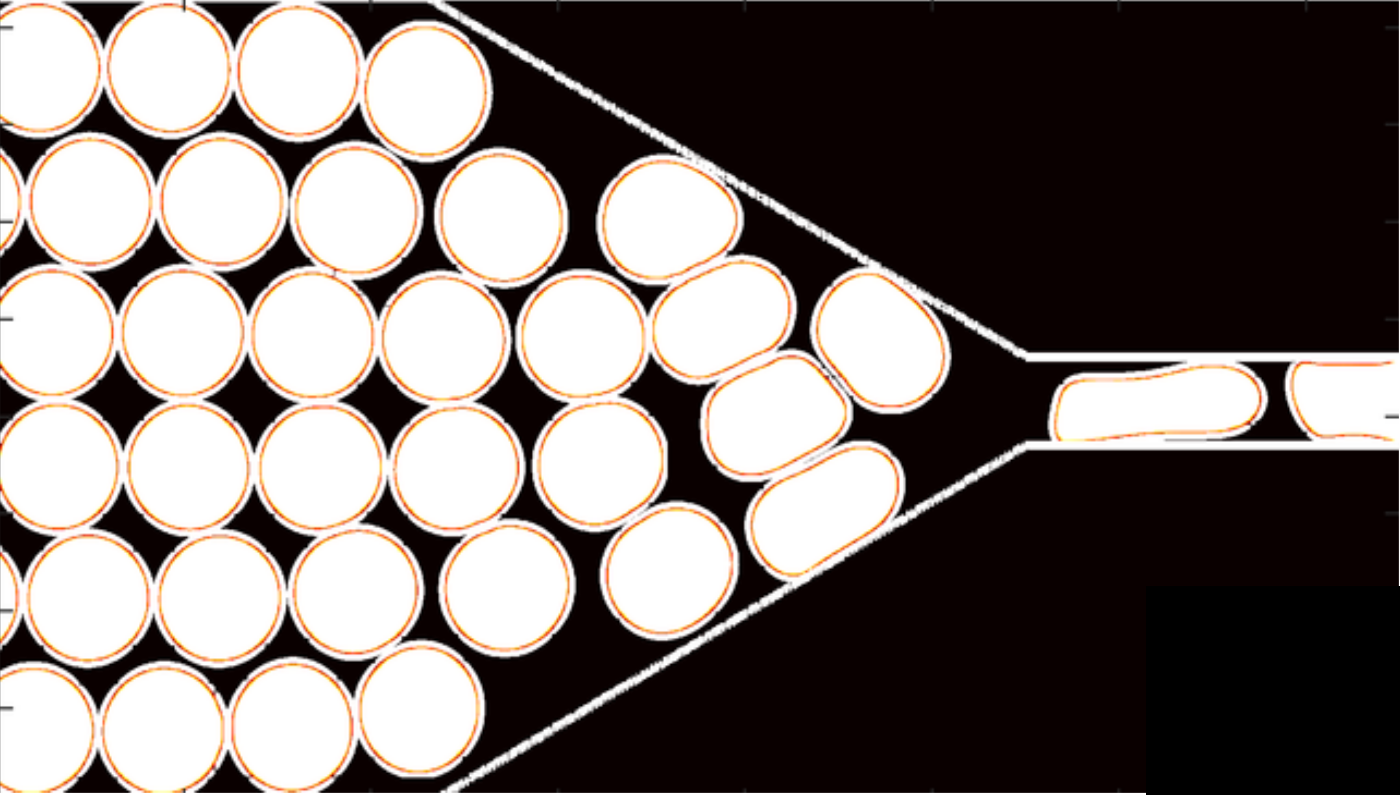}
  \caption{\label{Fig_tapered} A snapshot of the Tapered channel. Droplets enter the convergent nozzle from the left, pushing to enter the end of the channel at the right. The motion is now fully two-dimensional, with significant droplet deformations, thus setting
  a more demanding target to the Deep Learning procedure.}
\end{figure*}

\begin{figure*}[h]
  \centering
  \includegraphics[width=\textwidth,height=5cm,keepaspectratio]{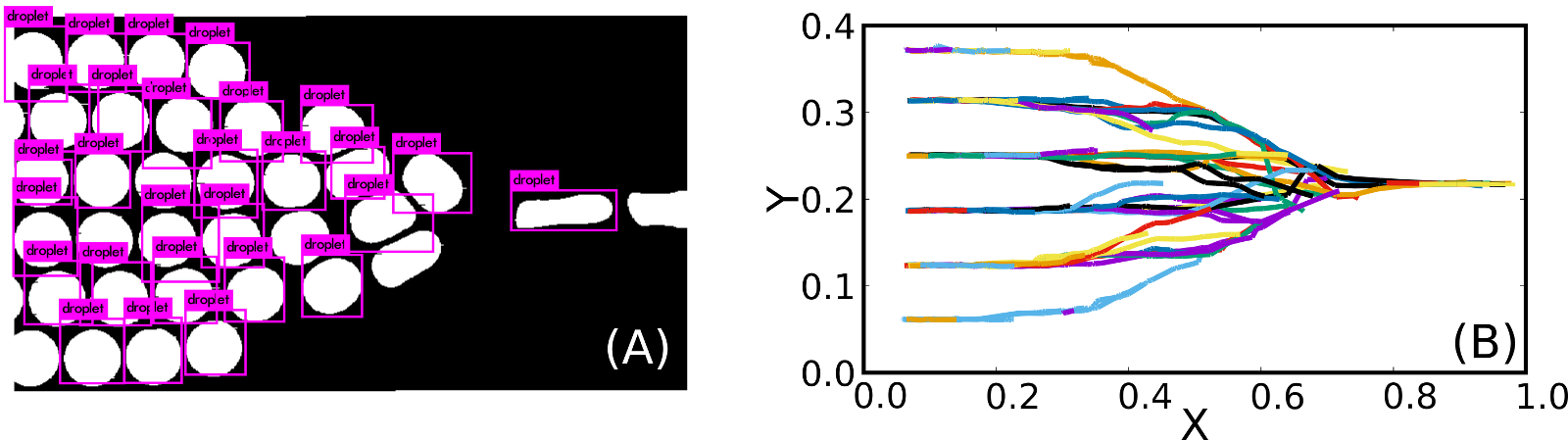}
  \caption{\label{traj_tapered} (A) Droplet detected for the Tapered channel by YOLO network for the same snapshot shown in Fig. \ref{Fig_tapered}. (B) DeepSort uniquely labels each droplet to compute all the trajectories. }
\end{figure*}

The second system (\textbf{Tapered channel}) consists of a dense emulsion, flowing inside a tapered channel connected to a thin straight channel (see Fig.~\ref{Fig_tapered}). Simulation output images of size 950 X 500 pixels were stacked to prepare input for the YOLO network. The predicted bounding boxes are shown in Fig. \ref{traj_tapered}A and the trajectories of the droplets, i.e. the centers of the predicted bounding boxes, are compiled by analysing all the frames are shown in Fig. \ref{traj_tapered}B. A video file (tapered\_channel.mp4) of the detection and tracking process is also provided in the Supplementary Material. 

\subsection{Computational performance}

The computational time required to identify and track objects is an important factor for many 
applications. In Table~\ref{inference_table}, we report the performances relative to the full track-and-detect process on different computers, measuring the inference speed in frames per second ({\it fps}).
We report the speed on two machines using only general purpose CPUs (in a typical configuration for a notebook and for a computing server), and using a middle level GPU for a desktop (NVIDIA RTX 2060 Super).
From this table, it is clear that the inference proceeds at the significant speed of 400 frames per second on the machine powered by GPUs. 

In another approach, feature extraction and clustering algorithms have been used to track moving objects at around 35fps \cite{keivani} on multi-core CPUs. Even though it is not fair to compare the inference speed for algorithms with data sets of different complexity and using different computing machines, it is worthwhile to note that, with the method demonstrated in this work, the complete process of identification and tracking of droplets can take place at much higher rates (few hundred {\it fps}) without trading-off the accuracy, and 
taking advantage of the existing machine learning frameworks highly optimized for the GPUs.

\begin{table}[!h]
    \centering
    \begin{tabular}{|| c | c | c ||}
       \hline
       \multicolumn{1}{||c|}{\textbf{Machine}}
       &\multicolumn{2}{c||}{\textbf{Inference speed}}\\
       \cline{2-3}
    & \textbf{Simple channel} & \textbf{Tapered channel} \\
 \hline\hline
       CPU ( 1X Intel i3-2328M)  &  1.61 fps & 0.69 fps \\
       CPU (16X Intel Xeon Platinum 8176) & 8.14 fps & 5.10  fps\\
       GPU (NVIDIA RTX 2060 Super) & 400 fps & 370 fps \\
        \hline
    \end{tabular}
    \caption{Average inference speed for the two cases on various types of devices in frame per second (fps).
    Performances on GPU are almost 100x the values obtained with a full CPU implementation. }
    \label{inference_table}
\end{table}

The application developed by training a cutting-edge deep learning model can be efficiently used 
to compute the trajectories of droplets in dense emulsion systems simulated by Lattice Boltzmann 
method or any other computational method capable of capturing the relevant physics. 
It is worth noting that in both cases explored, and especially in the case of the Tapered channel, the 
data used to train the network depart noticeably from those acquired via simulations. 
The fact that the network is nevertheless able to identify the droplets and efficiently 
track them over sequential frames bodes well for its generalisation to more complex
applications, particularly for tracking deformable biological bodies in complex fluid flows.
This will make the subject of future work.

\section{Conclusions}
In this work, we demonstrated an automated procedure to adapt a cutting edge deep learning-based 
application, called YOLO and DeepSORT, to infer trajectories of droplets in dense 
emulsions in microchannels. 
We have shown that the deep network can effectively be trained with synthetic data,  thereby 
bypassing the labour intensive process of acquiring the training data.
The present tool performs competitively with state of the art object tracking 
and detection algorithms in terms of inference speed. 

 Extracting the information about the trajectories could also provide useful insights in the study of several 
 other complex many-body systems, such as moving groups of animals and biological microorganisms. 
We also plan to extend the ML procedure not only to detect droplets 
and track their trajectories, but also to exploit the latter data to infer the effective 
equations of motion of this complex many-body system, possibly using generative 
adversarial networks \cite{yang2020}.\vskip6pt

\enlargethispage{20pt}






\section{Acknowledgement}{The research leading to these results has received funding from the European Research
Council under the Horizon 2020 Programme Grant Agreement n. 739964 ("COPMAT").
A.M. acknowledges the CINECA Computational Grant ISCRA-C IsC83 - “SDROMOL”, id. HP10CZXK6R under the ISCRA initiative, for 
the availability of high performance computing resources and support.}



%

\end{document}